\documentclass[twocolumn,showpacs,amsmath,amssymb,superscriptaddress]{revtex4}
\usepackage{color,longtable}
\usepackage[dvips]{graphicx}
\usepackage{amssymb,amsfonts,amsmath}
\begin{document}

\title{Measuring every particle's size from three-dimensional
imaging experiments}
\author{Rei Kurita} 
\affiliation{Institute of Industrial Science, 
The University of Tokyo, 4-6-1 Komaba, Meguro-ku, Tokyo 153-8505, Japan}
\email{kurita0@iis.u-tokyo.ac.jp}
\author{Eric R. Weeks}
\affiliation{Department of Physics, Emory University, 
Atlanta, Georgia 30322, USA}

\date{\today}

\begin{abstract}
Often experimentalists study particulate samples that are nominally
monodisperse.  In reality many samples have a polydispersity of
4-10\%.  At the level of an individual particle, the consequences
of this polydispersity are unknown as it is difficult to measure
an individual particle size from images of a dense sample.
We propose a method to estimate individual particle radii from
three-dimensional data of the particle positions.  We validate our
method by numerical simulations of four major systems: random close
packing, colloidal gels, nominally monodisperse dense samples,
and nominally binary dense samples.  We then apply our method
to experimental data from moderately concentrated colloidal
suspensions observed with confocal microscopy.  We demonstrate
that we can recover the full particle size distribution {\it
in situ}.  Lastly, we use our method to study the relationship
between homogeneous colloidal crystal nucleation and particle sizes.
We show that nucleation occurs in regions that are more monodisperse
than average.
\end{abstract}

\maketitle

A wide variety of techniques exist for three-dimensional imaging
of collections of particles \cite{dijksman12,prasad07}.  These
types of samples include granular materials, soil mechanics,
and colloidal suspensions.  Our particular interest is in
colloidal suspensions; these have been successfully used as
model systems for understanding phase transitions for several
decades \cite{Pusey,Lekkerkerker}, and moreover are
interesting in their own right due to industrial relevance
\cite{Vincent}.  Confocal microscopy can be used to take
three-dimensional images of fluorescent colloidal particles
deep within a sample \cite{vanblaaderen92,Blaaderen1,prasad07}.
When coupled with particle tracking techniques, the motion of
thousands of individual colloidal particles can be followed over
long periods of time \cite{kegel00,EricSci,Eric2002,Dinsmore}.
This technique has been used to investigate the colloidal glass
transition \cite{Blaaderen1,kegel00,EricSci,Eric2002,Narumi},
crystallization \cite{Blaaderen2,Gasser,Dullens06}, colloidal
gels \cite{dinsmore02gel,dibble06,gao07}, capillary waves
\cite{aarts04,hernandez09}, sedimentation \cite{Blaaderen2,Paddy},
and a variety of other questions (see ref.~\cite{prasad07} for a
review).  One advantage of confocal microscopy of colloids is that
the data obtained are similar to what is found using simulations,
which also provide the data of particle positions over long periods
of time.

However, experimental samples are always polydisperse:  even
for a nominally single-component sample, the particles have
a variety of sizes \cite{poon12}.  This is quantified by the
polydispersity $p$, defined as the standard deviation of particle
radii divided by the mean radius.  For many samples, $p \sim 0.04
- 0.10$ \cite{poon12}.  From numerical simulations, we know that
the effects of the particle size distribution are not negligible.
For example, crystal nucleation is difficult or impossible for more
polydisperse samples \cite{Frenkel,pusey09}.  The crystal-liquid
phase boundary depends on the polydispersity \cite{sollich10}.
The sensitivity to volume fraction near the glass transition
depends on the composition in nontrivial ways \cite{KAT,KW2}.
Experimentally, the influence of polydispersity on colloidal
crystallization has been demonstrated \cite{schope07,henderson98},
and there is also some understanding of how the particle size
distribution influences the rheological behavior of a colloidal
sample \cite{mewis94}.  However, these are limited to studies of the
spatially averaged properties of the sample.  Microscopy is
useful for local properties, but 
particle size fluctuations of $0.04-0.10$ are not easily
detectable.
It would be
desirable to know particle sizes for more direct comparison with
simulations.  Furthermore, in some cases, neglecting these sizes
in an experiment can lead to wrong conclusions.  One example is
that the pair correlation function $g(r)$ can show a qualitatively
incorrect dependence on control parameters if the particle sizes
are treated as all identical \cite{pond11}.  A second example is
that the apparent compressibility of a random close packed sample
depends qualitatively on whether individual particle sizes are
taken into account \cite{Berthier,Zachary1,KW3}.

In this work, we introduce a general method for using 3D data
to determine the size of individual particles in any moderately
concentrated sample, in general with volume fractions $\phi
\gtrsim 0.4$.  We use simulation data to verify that our method
works well in a variety of sample types.  We then demonstrate the
utility of our method using previously published experimental
data from confocal microscopy of colloids.  In particular, we
show that colloidal crystal nucleation is sensitive to the local
polydispersity: nucleation happens in locally monodisperse regions.
Our method is not limited to confocal microscopy and colloidal
samples, but rather works with any data of the 3D positions of a
collection of particles.

Due to diffraction limits, it is difficult to directly determine
the radii of individual particles from microscopy images to better
than $\pm 0.1$~$\mu$m \cite{brujic03}.  Defining the edge is
somewhat arbitrary and varies depending on particle properties and
the details of the microscope illumination.
Other 3D imaging techniques have similar
issues \cite{dijksman12}.

In contrast, it is much easier to calculate the mean radius
$\bar{a}$ of particles with a variety of techniques \cite{poon12}.
Likewise, from the centers of particles, the separations between
neighboring particles $r_{ij}$ can be easily calculated.  
Our estimation method for particle sizes uses only $\bar{a}$
and $r_{ij}$.  The key idea of our method is that
a large particle will be slightly farther from its neighbors and
thus have larger values for its $r_{ij}$, and likewise a smaller
particle will have smaller values of $r_{ij}$.

To start, we relate the pairwise separations $r_{ij}$ as
\begin{equation}
r_{ij}(t) = a_i + a_j + \delta_{ij}(t),
\end{equation}
where particle $j$ is a nearest neighbor particle of particle $i$,
$r_{ij}$ is the measured distance between $i$ and $j$, $a_i$ and
$a_j$ are their radii, and $\delta_{ij}(t)$ is a surface-to-surface
distance between their particles.  We typically consider $5 - 7$
nearest neighbor particles (the closest neighbors);
this choice is justified below.
Often
these data come from particle tracking \cite{Dinsmore,Crocker}
and so $r_{ij}(t)$ and $\delta_{ij}(t)$ depend on time $t$.
Next we take an average of
$r_{ij}$ with respect to the nearest neighbor particles $j$, and
then $\langle r_{ij}(t) \rangle_{j} = a_i + \langle a_j \rangle_{j}
+ \langle \delta_{ij}(t) \rangle_j$, where $\langle \rangle_{j}$
means an average over particle $j$.  Thus, we obtain
\begin{equation}
a_i = \langle r_{ij}(t) \rangle_{j} - \langle a_j \rangle_{j} - \langle \delta_{ij}(t) \rangle_j \label{eq:1}.
\end{equation}
This is exact, but the quantities $\delta_{ij}(t)$ are unknown.
We estimate this by replacing $\delta_{ij}(t)$ with its time- and
particle-averaged value, the mean gap distance $\bar{\delta} \equiv \langle
r_{ij}(t) \rangle_{i,j,t} - 2\bar{a}$, where the average is over
all particle pairs and all times.  Our algorithm is then:
\begin{eqnarray}
a_i^{(0)}(t) &=& \bar{a} \label{eq:2}, \\
a_i^{(n)}(t) &=& \langle r_{ij}(t) \rangle_{j} - \langle a_j^{(n-1)}(t) \rangle_{j} - \bar{\delta}, \label{eq:3} 
\end{eqnarray}
where the superscripts denote iteration.  The more
we iterate Eq.~\ref{eq:3}, the more information we obtain from
particles far away from a given particle.  In fact, $a_i^{(n)}(t)$
is unchanged for $n \ge 10$ since $a_i^{(10)}$ includes the
information from several thousand particles, thus we fix $n=10$ for
the number of iterations in this paper.  Of course, the particle
radius does not depend on time, so after the 10th iteration,
we time-average $a_i^{(10)}(t)$ to obtain the estimated particle
radius $a_i^{(10)}$.  Time-averaging after each iteration of
Eq.~\ref{eq:3} does not change the results.  

There are several sources of uncertainty in this estimation.  First,
there is the uncertainty of each particle position.
Typically this is about 5-8\% of the mean radius, leading to a
8-10\% uncertainty of $r_{ij}$ \cite{EricSci,Dinsmore,Crocker}.
However, these errors are nearly time-independent, so
those errors are greatly diminished by time averaging.  Second,
our approximation for $\bar{\delta}$ is weaker in the case
that the distribution of $\delta_{ij}(t)$ is broad.  This in
part depends on how many nearest neighbor particles are chosen:
more neighbors results in a broader distribution, whereas too
few neighbors means that the average $\langle r_{ij}(t)\rangle$
in Eq.~\ref{eq:3} is poor.  Below, we use simulation data to
determine that $Z = 5 - 7$ nearest neighbors is an optimal choice.
Third, independent of a given choice of $Z$, some particles will
simply be farther from their neighbors, and some will be closer.
In a dense suspension, for example, this relates to the size of the
``cage'' formed by the nearest neighbor particles \cite{Eric2002}.
Again, time averaging helps.  If particles can rearrange
and find new neighbors, then $\bar{\delta}$ becomes a better
approximation for $\langle \delta_{ij}(t) \rangle_t$.  In dense
colloidal suspensions with volume fractions $\phi \gtrsim 0.5$,
rearrangements become infrequent and so longer time averages are
desired \cite{kegel00,EricSci,Eric2002}.  In summary, the greatest
strength of our algorithm is time-averaging, and past that, a
sensible choice for the number of nearest neighbors $Z$ is useful.
Our tests show that time averaging over $\sim 20$ different times
is sufficient for reasonable results.

To verify our radius estimation method, we simulate a
variety of systems and compare the estimated radius of
each particle with its true radius.  The error is given by $\delta
a_i = a_i^{(10)}-a_i$, where $a_i^{(10)}$ is the estimated value
and $a_i$ is the true value.  $\Delta a \equiv \langle \delta a_i
\rangle / \bar{a}$ is the mean fractional error in the estimated
particle radius.  Also relevant is the polydispersity $p$ of the
simulated sample, defined as $p=\sqrt{\langle (a_i - \bar{a})^2
\rangle}/\bar{a}$, where the averages are over all
particles $i$.  Before any estimation is applied, the best guess for
each particle size is $\bar{a}$ with a fractional
uncertainty $p$.  If the mean estimation error $\Delta a$ is less
than $p$, the estimation method improves our knowledge of the
particle sizes; we will show this is true for the simulated data.

\begin{figure}[t]
\begin{center}
\includegraphics[width=8cm]{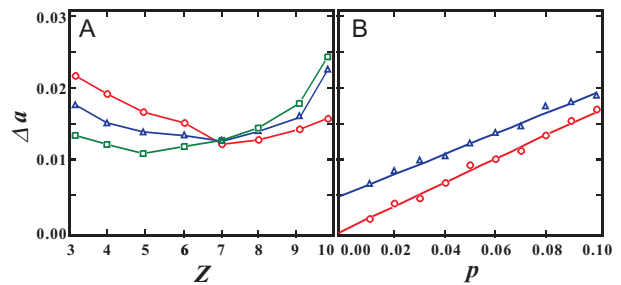}
\end{center}
\caption{(Color online) Dependence of estimation uncertainty
$\Delta a$ on parameters.  (A) The uncertainty 
$\Delta a$ as a function of the chosen number of neighbors $Z$
used for the averaging.  The circles, triangles and squares
correspond to $\Delta a$ for volume fractions $\phi$ = 0.51,
0.56, and 0.64 (RCP), respectively.  In each case, the sample
polydispersity is $p=0.07$.  (B) $\Delta a$ as a function of the
bulk polydispersity $p$. The circles and the triangles correspond
to $\Delta a$ at RCP and $\phi$ = 0.51, respectively. The solid
lines are the fitting lines for $\Delta a$.
\label{Z-delta}
}
\end{figure}

Our first test case is a random close packed sample.  In such a
sample particles do not move, and so time-averaging cannot be used.
However, particles are packed so that they contact each other,
that is, $\bar{\delta} = 0$.  The number of contacting neighbors
varies from particle to particle, so it is not clear how many
neighbors should be considered.  Accordingly, we plot $\Delta a$
as a function of $Z$ in Fig.~\ref{Z-delta}{\it A}.  We find that
$\Delta a$ is a minimum at $Z$ = 5, and is indeed much smaller
than $p$ (0.01 {\it vs.} 0.07 in this case).

\begin{figure}[t]
\begin{center}
\includegraphics[width=8cm]{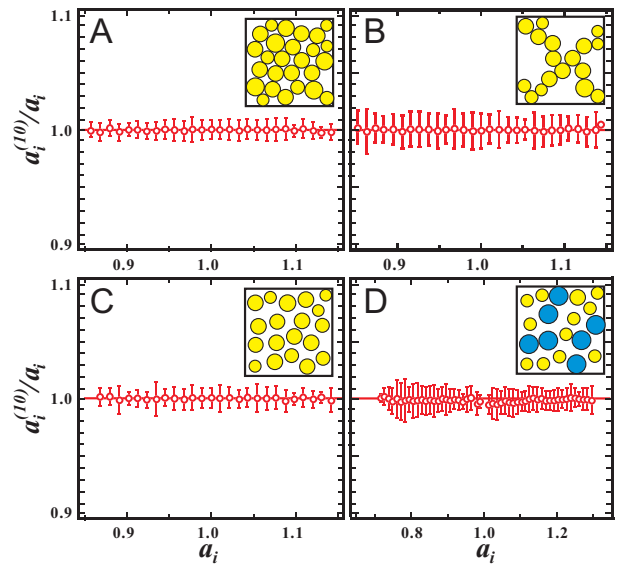}
\end{center}
\caption{(Color online) The estimated radius $a_i^{(10)}/a_i$ as
a function of the true radius $a_i$, for four simulated systems:
(A) random close packing, (B) a colloidal gel at $\phi = 0.10$,
(C) a nominally monodisperse suspension at $\phi = 0.51$, and
(D) a nominally  binary suspension at $\phi = 0.51$.  For (A-C),
the polydispersities are $p=0.07$.  For the binary sample, the size
ratio is $1:1.3$, the number ratio is $1:1$, and each species has an
individual polydispersity of $p=0.04$.  The error bars correspond
to the standard deviation of of $a_i^{(10)}/a_i$ between $a_i$
and $a_i+0.01$.  The insets show sketches of each system.
\label{size-est}
}
\end{figure}

It is possible that while $\Delta a$ is small, that there are
systematic errors depending on the real particle size $a_i$.
To test this, in Fig.~\ref{size-est}{\it A} we show the ratio between
the estimated radius and the given radius $a_i^{(10)}/a_i$
as a function of $a_i$ for a polydispersity $p=0.07$ RCP sample.
The symbols and the error bars correspond to the mean
and standard deviation of the distribution of $a_i^{(10)}/a_i$
between [$a_i, a_i+0.01$], respectively.  $a_i^{(10)}/a_i$ should
be 1 if our estimation is perfect and indeed we
find $a_i^{(10)}/a_i = 1.000 \pm 0.013$.  The quality of the
results is nearly uniform as a function of particle size.
To check the validity of our method for RCP samples with different
polydispersity, we plot the uncertainty $\Delta a$ as a function
of sample polydispersity $p$ in Fig.~\ref{Z-delta}{\it B}.  We find
$\Delta a \approx p/6$ \cite{KW3}.

A colloidal gel shares a similarity to a RCP sample (touching
particles), and has a significant difference (much lower
volume fraction).  In a colloidal gel particles are stuck to
their neighbors and form a large network.  Often the attractive
interactions have a finite range, for example with depletion gels
\cite{AO} [see discussion in {\it Methods}].  Thus we
note that the distribution of $\delta_{ij}$ for gels is slightly
broader than that for RCP, though the mean average of $\delta_{ij}$
is close to 0.  Some time averaging is possible, although such
samples are frequently nonergodic or at best rearrange quite slowly.

Likewise the contacting particles make gels similar to
RCP samples locally.  However, the contact number fluctuates greatly
in a colloidal gel, and the number of neighbors averaged over
must vary from particle to particle.  Rather than
being a fixed parameter $Z$, we have a varying number of neighbors
$Z_i$ used in the average (Eq.~\ref{eq:3}).  To determine $Z_i$,
we define the coordination number $c_i$ as the number of 
particles within a distance 2.8$a$, which is the first minimum of
the pair correlation function.  We find the average coordination
number $\bar{c} \approx 13.1$ for a RCP sample, but this will
generally be smaller for a gel \cite{dinsmore02gel}.  Thus for
every particle we estimate the number of touching neighbors $Z_i =
5c_i/13$ where we round $Z_i$ to the nearest integer.  In general,
given the tenuous nature of a gel, for many particles $Z_i$ is
fairly small; also, $\delta_{ij}$ has a broader distribution, and
so $\Delta a$ will be worse than the RCP case.  However, $\Delta a$
is improved by time-averaging, which also minimizes the uncertainty
due to particle tracking errors.  Fig.~\ref{size-est}{\it B}
shows the ratio between the time-averaged estimated radius and the
given radius $a_i^{(10)}/a_i$ as a function of the true radius
$a_i$ for the colloidal gel.  We find that $a_i^{(10)}/a_i =
1.000 \pm 0.018$.  $\Delta a = 0.018$ is much smaller than
the polydispersity $p=0.07$.

Moving from gels, we next consider a dense suspension of purely
repulsive (hard-sphere) particles.  Here no particles are in
contact, so $\delta_{ij}$ has a much broader distribution; however,
time-averaging is even more powerful.  We show $a_i^{(10)}/a_i$ as
a function of $a_i$ at $\phi = 0.51$ in Fig.~\ref{size-est}{\it C},
finding $a_i^{(10)}/a_i = 1.000 \pm 0.014$.  Yet again $\Delta a =
0.014$ is much smaller than the polydispersity $p=0.070$.

For a dense suspension it is not obvious how many nearest neighbors
should be used in the average (Eq.~\ref{eq:3}), so we plot $\Delta
a$ as a function of $Z$ in Fig.~\ref{Z-delta}{\it A} for two
different volume fractions.  $\Delta a$ is minimized at $Z = 7$
for the non-RCP samples (circles and triangles in the figure), so
we fix our choice $Z=7$ for all our $\phi < 0.6$ experimental data
(discussed below).  Figure~\ref{Z-delta}{\it A} demonstrates that
$\Delta a$ does not depend too sensitively on this choice.  However,
it should be expected that for a more dilute system, the importance
of caging decreases, and the number of neighbors a particle has
will fluctuate significantly.  For fixed polydispersity $p=0.070$,
we find $\Delta a = 0.023$ for $\phi=0.45$ and $\Delta a = 0.060$
for $\phi=0.40$.  This suggests that for $\phi \lesssim 0.40$, the
estimation method may not be useful without further modifications.

To check the influence of the sample polydispersity
at fixed $\phi=0.51$, we vary $p$
with results shown in Fig.~\ref{Z-delta}{\it B}
(triangles).  We find $\Delta a \approx 0.005 + p/7$, suggesting
that the estimation is useful for samples with $p> 0.01$, that is,
any realistic sample.  $\Delta a$ is nonzero when $p=0$,
in contrast to the RCP case.  This is due to the distribution of
$\delta_{ij}$ in a dense but non-contacting sample.

The last case we examine with simulation data is a nominally
binary sample.  We simulate a dense suspension composed of
particles with a size ratio 1:1.3 (mean sizes 0.877 and 1.14) and
number ratio $1:1$.  For both ``small'' and ``large'' particles,
there is a polydispersity $p=0.04$.  The results are shown in
Fig.~\ref{size-est}{\it D}, and we find $a_i^{(10)}/a_i = 1.000 \pm
0.024$ at $\phi = 0.51$.  (Here we have fixed $Z=7$.)  Again,
there is no strong dependence on the true particle size $a_i$, and
in particular the particles in the tails of the distributions are
estimated with good accuracy.  However, the uncertainty $\Delta
a$ for the binary mixture is larger than what is found for the
nominally monodisperse distribution.  This is consistent with the
overall polydispersity of the sample being larger, $p=0.14$.

\begin{figure}[t]
\begin{center}
\includegraphics[width=8cm]{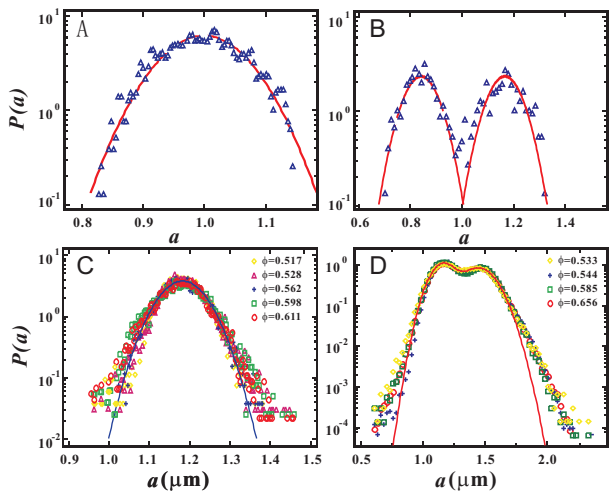}
\end{center}
\caption{(color online) Particle size distributions.  (A) The
distribution of the true radius (solid lines) and the estimated
radius $a_i^{(10)}$ (triangles) for a simulated nominally
monodisperse sample with $\phi=0.51$ and polydispersity $p=0.07$.
(B) The distribution of the true radius (solid lines) and the
estimated radius $a_i^{(10)}$ (triangles) for a simulated nominally
binary sample with $\phi=0.51$; see text for more details.
(C) The radii distributions for the nominally monodisperse experimental
suspension from ref.~\cite{EricSci}, for five volume fractions as
indicated.  The solid line is a Gaussian fit to the combined data,
giving $p=0.045$.  (D) The radii distributions for the nominally
binary experimental suspension from ref.~\cite{Narumi}, for four
volume fractions as indicated.  Here the solid line is a fit to the
sum of two Gaussians.  With the size ratio (1:1.3) and
the polydispersity of each species ($p=0.049$ for the small species,
$p=0.050$ for the large species), the two sub-distributions have
substantial overlap.
\label{distribution}
}
\end{figure}

An important use of the estimation technique is to measure the
particle size distribution of a sample {\it in situ}; we wish to
validate this idea with the simulation data.  To do this, we compare
the estimated radius distribution $P(a_i^{(10)})$ with the true
radius distribution $P(a_i)$ in Fig.~\ref{distribution}{\it A,B}.
In both the nominally monodisperse sample and the nominally binary
sample, the estimated distribution (symbols) is quite close
to the true distribution (lines).  Our results show that the
estimated distribution is essentially reproduced by convolving
the true distribution with a Gaussian of width $\Delta a$.
For a single-species sample with a Gaussian distribution of radii
with polydispersity $p$, the estimated polydispersity would be
$\sqrt{p^2+\Delta a^2}$.  Given that for most situations we have
shown $\Delta a \ll p$, our technique will only slightly increase
the apparent polydispersity of a sample.

One key difference between simulations and experiments is the
boundary condition.  Our simulations have periodic boundaries.
In an experiment, we can not find all nearest neighbors of a
particle when the particle is located at the edges of an image.
This situation is similar to colloidal gels, where the number
of nearest neighbors varies for each particle, and we adopt the
same solution used there.  For each particle, we average over a
number of nearest neighbors given by $Z_i = 7c_i/13$, where $c_i$
is the observed coordination number defined before, and we round
$Z_i$ to the nearest integer.  The denominator 13 is chosen as the
number of neighbors in a close-packed sample, and the numerator
7 is from the results of Fig.~\ref{Z-delta}{\it A}.

Furthermore, we need one more improvement when we apply our
method to a nominally binary sample.  It usually happens that
we know the mean radii of each of the two species, while the
number ratio of two species is unknown, which means that $\bar{a}$
is unknown.  In this situation, we start with a reasonable guess
for $\bar{a}'$ to be used in Eq.~\ref{eq:2}.  Then we compute
the particle radii and obtain the double peak distribution which
depends on our guess $\bar{a}'$.  Both peak radii of the trial
estimated radius distribution should be shifted by $(\bar{a}' -
\bar{a})$ from the known mean radii.  Thus we subtract $(\bar{a}'
- \bar{a})$ to adjust the peak positions to the known mean radius
of each species and we obtain the estimated particle size.

In an experiment we do not have an alternate means to determine
each particle size and so cannot directly verify our results
in the way that the simulations allow.  However, evidence that
our method works is shown in Fig.~\ref{distribution}{\it C,D}.
Here, we analyzed previously published experimental data from
ref.~\cite{EricSci} (nominally monodisperse) and ref.~\cite{Narumi}
(nominally binary).  In each case, data from several different
volume fractions are shown.  The size distributions agree well
for the different volume fractions for both the monodisperse and
binary cases.  Each different volume fraction was a sample taken
from the same stock jar and therefore should have the same size
distribution, so this is a confirmation that our method works well
with experimental data.

We now demonstrate the utility of our algorithm by studying
colloidal crystal nucleation.  The nucleation of crystals in a
dense particle suspension depends sensitively on polydispersity
\cite{pusey09,schope07,henderson98}.  We examine data of the
$\phi=0.46$ sample from ref.~\cite{Eric2002}, analyzed at longer
times to examine the crystallization process that was discarded
from the analysis in ref.~\cite{Eric2002}.  These particles are
slightly charged, shifting the freezing point to $\phi_{\rm freeze}
\approx 0.38$ and the melting point to $\phi_{\rm melt} \approx
0.42$ \cite{Gasser}.  In this data, we confirm that the crystal
nucleus appears at the center of our microscopic image: this is
homogeneous nucleation, not heterogeneous nucleation near the wall.

\begin{figure}[t]
\begin{center}
\includegraphics[width=8cm]{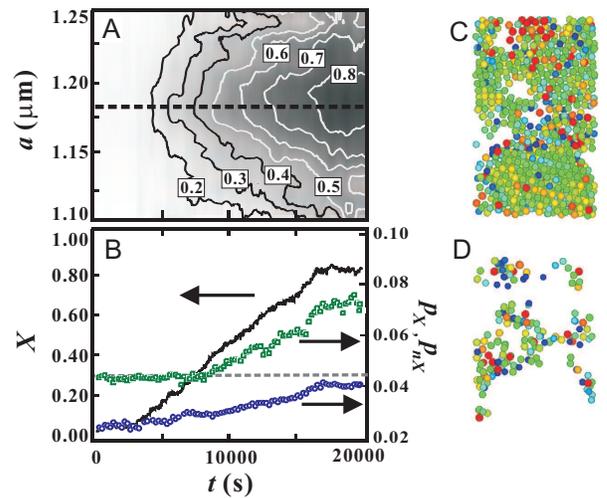}
\end{center}
\caption{Relationship between individual particle size $a_i$ and
crystallization.  ({\it A}) The contour plot of the fraction 
of crystalline particles $X$ as a function of $a_i$ and time $t$.
The contour lines are numbered by the value of $X$.
The dashed line shows the mean radius $\bar{a}$. 
Particles with radii close to $\bar{a}$ crystallize faster.
({\it B}) The fraction of the sample $X$ that is crystalline as
a function of time (solid black line), along with the mean 
polydispersity of all crystalline particles (blue
circles) and all non-crystalline particles (green squares).
The gray dashed line corresponds to the bulk polydispersity
$p=0.045$.
({\it C}) The crystalline particles are shown at $t = 20000$~s.
({\it D})  The non-crystalline particles are shown at $t=20000$~s.
For {\it C} and {\it D}, the color indicates $a_i$ for each
particle, where blue corresponds to smaller $a_i$, light green corresponds 
to $a_i$ close to $\bar{a}$, and red corresponds to larger $a_i$.
\label{Xtal}
}
\end{figure}

At each time step, we calculate the number of ordered
neighbors $N_o$ for each particle using standard techniques
\cite{Gasser,Wolde} [see {\it Methods}].  By convention,
a crystalline particle has $N_o \geq 8$ \cite{Gasser,Wolde}.
At each time we compute the number fraction of the sample that
is crystallized, $X(t)$.  Figure~\ref{Xtal}{\it A} shows $X(t)$
as a function of both individual particle size and time, where
darker colors correspond to larger values of $X(t)$.  Below $t =
3000$~s, $X < 0.2$ for all $a$, and essentially all crystal clusters
are below the critical size ($\sim 100$ particles) \cite{Gasser}.
At $t = 3000$~s, a sufficiently large crystalline region appears
and begins to grow.  $X$ increases first for particles with $a$
close to the mean radius, and these particles continue to be the
subpopulation that is the most crystallized at any given time.
At longer times the particles with $a$ farther from $\bar{a}$
gradually begin to crystallize.

We next consider an alternate way of thinking about the same data.
Figure~\ref{Xtal}{\it B} shows the relationship between the
sample-averaged $X(t)$ (solid black line), the polydispersity $p_X$
for all crystalline particles (blue circles), and the polydispersity
$p_{nX}$ for all non-crystalline particles (green squares).  $X$
starts to increase at $t = 3000$~s, and those particles that are
crystalline at that time have $p_X \sim 0.03$, smaller than the
bulk polydispersity $p=0.045$.  As the sample crystallizes we
observe that both $p_X$ and $p_{nX}$ increase.  The growth of
$p_X$ indicates that the crystal, while nucleating in a fairly
monodisperse region, can grow by incorporating particles that
are farther from the mean size.  In the final state, the local
polydispersity of the crystalline particles has nearly reached
the mean polydispersity $p$.  The growth of $p_{nX}$ indicates
that those particles that are still outside the crystal are more
likely to be those with unusual sizes.

The spatial distribution of particles at the end of the experiment
is shown in Fig.~\ref{Xtal}{\it C,D}.  Figure \ref{Xtal}{\it
C} shows the locations of the crystalline particles, while
{\it D} shows the locations of the non-crystalline particles.
Green particles have $a_i$ close to $\bar{a}$, while the smallest
particles are drawn blue and the largest drawn red.  The cores of
the crystal regions are composed of the green particles.

\begin{figure}[t]
\begin{center}
\includegraphics[width=6cm]{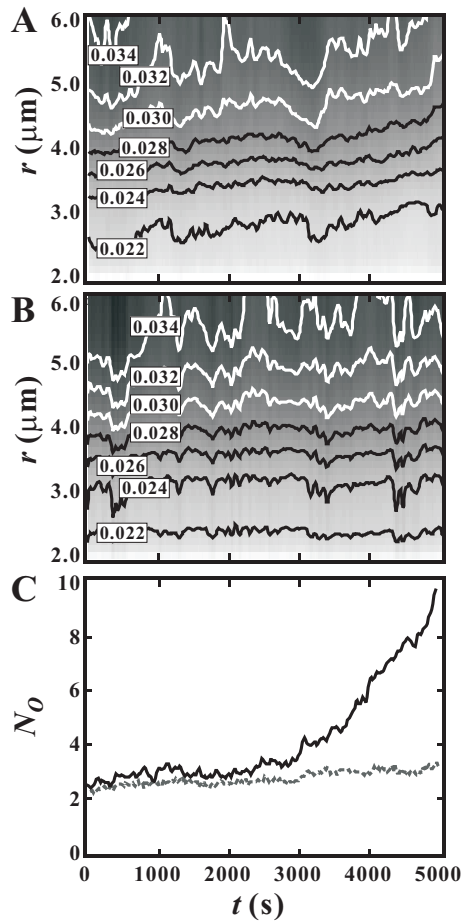}
\end{center}
\caption{
Relationship between nucleation properties
and the local polydispersity $p_i(r)$.
({\it A}) The contour plot of the mean $p_i(r)$
as a function of $r$ and $t$, averaged over all particles which
are crystalline at $t = 5000$~s.  The numbers on the contour
lines represent the value of $p_i(r)$.  ({\it B}) The contour
plot of the mean $p_i(r)$ as a function of $r$ and $t$, averaged
over all particles which are non-crystalline at $t=5000$~s.
The particles considered in ({\it A}) and ({\it B}) are only
those with radii close to the mean radii; see text for details.
({\it C}) The number of ordered neighbors $N_o$ as a function of
time for those particles plotted in ({\it A}) (nucleus particles,
solid line) and ({\it B}) (non-nucleus particles, dashed line).
\label{LocalP}
}
\end{figure}

Next, we examine the beginning of the crystal nucleation process.
While many particles are close
to the mean size, only a few end up being the nucleation site.
To understand which ones nucleate, we now focus
on the particles close to the mean size: radii 1.175 $\mu$m
$< a_i < $ 1.185 $\mu$m.  Among those particles, we define the
nucleus particles as those that are crystalline particles at $t =
5000$~s; the remainder are non-nucleus particles.  We next
define the {\it local} polydispersity $p_i(r)$ of particle $i$ as
\begin{equation}
p_i(r) = \sqrt{\langle (a_j - a_i)^2 \rangle}/a_i
\label{localpoly}
\end{equation}
where the angle brackets $\langle \rangle$ indicate an average over
all particles $j$ with centers within a distance $r$ from particle
$i$.  Figures~\ref{LocalP}{\it A,B} show space-time plots of the mean
value of $p_i(r)$ for the nucleus particles {\it (A)} and
the non-nucleus particles {\it (B)}.  
In all cases, $p_i(r)$
is lower close to the particles and increases with increasing $r$.
However, notably the contours for low $p_i$ are at smaller values
of $r$ for the nucleus particles {\it (A)}.  After $t \approx
3000$~s, the region of low $p_i$ spreads to large values of $r$
for the nucleus particles {\it (A)}, while little change is seen
for the non-nucleus particles {\it (B)}.  Simultaneously, we show
the temporal change of $N_o$ in Fig.~\ref{LocalP}{\it C} for the
nucleus particles (solid black line) and the non-nucleus particles
(dashed gray line).  This confirms that the onset of crystallization
at $t \approx 3000$~s coincides with the expansion of the low
local polydispersity region seen in Fig.~\ref{LocalP}{\it A}.
This is all evidence that crystal nuclei are formed from regions
where the particles are all similar sizes.
A reasonable conjecture is that nucleation rates are possibly
quite sensitive to how well-mixed the sample initially is, in this
respect of local polydispersity.

We have developed a general method to estimate the particle sizes in
a dense particulate samples where the particle positions are known.
Simulations demonstrate the validity of our method.  This method can
be applied to any cases where three-dimensional particle positions
can be found; while we have focused on colloidal samples, granular
media are quite similar \cite{dijksman12,slotterback08}.  We have
demonstrated the utility of our method by examining homogeneous
colloidal crystal nucleation.  While it has been known that
nucleation is faster for more monodisperse samples, we find this is
true on a quite local scale.  Nucleation happens in regions that
are locally more monodisperse, and crystal growth is proceeds by
preferentially incorporating particles close to the mean size.

\section*{Materials and Methods}

We simulate four particle suspension systems, which are random
close packing (RCP), colloidal gel, single component suspension,
and a binary system.  The polydisperse RCP sample is generated using
the algorithm of ref.~\cite{Xu}.  For the three other cases, we
perform three-dimensional Monte Carlo simulations with hard spheres.
Additionally for gels, we wish to model colloid-polymer mixtures
and so we use the Asakura and Oosawa model \cite{AO}.  This model
leads to a pair interaction between two hard colloidal spheres in a
solution of ideal polymers as $U(r) = \infty$ for $r<\sigma_{ij}$,
$U(r) = -\frac{\pi}{12} k_BT \rho_{p} [r^3 -3(\sigma_{ij}+R_G)^2
r+2(\sigma_{ij}+R_G)^3]$ for $\sigma_{ij} \le r < \sigma_{ij} +
2R_G$, $U(r) = 0$ for $r \ge \sigma_{ij} +2R_G$, where $\sigma_{ij}
= (\sigma_i+\sigma_j)/2$, $\sigma_i$ is a diameter of particle
$i$, $k_B$ is Boltzmann constant, $T$ is temperature, $\rho_p$ is
the number density of polymers, and $R_G$ is the polymer radius of
gyration.  We fix $R_G = 0.1 \bar{\sigma}$ and $\phi_p = 4\pi/3
R_G^3 \rho_{P} = 0.1$ where $\bar{\sigma}$ is the mean diameter
of the hard spheres.  For our single-component and two-component
hard sphere suspensions, particles interact via $U(r) = \infty$
for $r<\sigma_{ij}$, otherwise $U(r) = 0$.  We use 1024 particles
with the mean radius $\bar{a}$ = 1 and variable polydispersity for
all simulations.

The experimental data come from prior experiments
\cite{EricSci,Narumi}.  These experiments used sterically
stabilized poly(methyl methacrylate) (PMMA) particles and imaged
them with confocal microscopy.  The particle positions were
located and tracked using standard particle tracking techniques
\cite{Dinsmore,Crocker}.  Detailed experimental discussions are
in the prior references.

We use previously developed order parameters to look for
crystalline particles and ordered structure
\cite{Gasser,Wolde,Steinhardt}.  For each particle $i$, we find
its nearest neighbors $j$ and identify unit vectors
$\hat{r}_{ij}$ pointing to the neighbors.  We then define a
complex order parameter $\hat{q}_{lm}$ using 
$q_{lm}(i)=\sum_{j=1}^{c_i} Y_{lm}(\hat{\bf r}_{ij})$ 
where $c_i$ is the number of nearest neighbors
of particle $i$ and $Y_{lm}$ is a spherical harmonic
function; we normalize this as $\hat{q}_{lm} = q_{lm}/N$ where
$N$ is a normalization factor such that
$\sum_{m}\hat{q}_{lm}(i)\hat{q}_{lm}^{*}(i)=1$ \cite{Gasser}.  We
use $l=6$.  For each particle pair, we compute the complex
inner product $d_6=\sum_{m}\hat{q}_{lm}(i) \hat{q}_{lm}^{*}(j)$.
Two neighboring particles are termed ``ordered neighbors" if $d_6$
exceeds a threshold value of 0.5.  For each particle, we focus on
$N_o$, the number of ordered neighbors it has at a given time.
$N_o^i$ measures the amount of similarity of structure around
neighboring particles. $N_o^i$ =0 corresponds to random structure
around particle $i$, while a large value of $N_o^i$ means that
particle $i$ and its neighbor particles have similar surroundings
\cite{Wolde}.

\section*{Acknowledgments}
E.~R.~W.~was supported by a grant from the National Science
Foundation (CHE-0910707).  We thank K.~Desmond and T.~Divoux for
helpful discussions.

\noindent
{\bf Competing interests statement} The authors declare that they 
have no competing financial interests.

\noindent
{\bf Correspondence} and requests for materials should be addressed to R. K.
(kurita0@iis.u-tokyo.ac.jp).

\end{document}